\documentclass[12pt,preprint]{aastex}

\usepackage{epsfig}
\usepackage{amsmath}
\usepackage{amssymb}
\usepackage{psfrag}
\begin{document}
\title{Erratum: ''Tidal Love numbers of neutron stars''\\
(2008, ApJ, 677, 1216) }

\author{Tanja Hinderer}
\affil{Center for Radiophysics and Space Research, Cornell
University, Ithaca, NY 14853} \email{tph25@caltech.edu}

There are typographical errors in Eqs. (20) and (23), and some
incorrect entries in Table (1). I thank Ryan Lang for pointing these
out.

Equation (20) should read as follows:
\begin{eqnarray}
H&=&c_1 \left({r \over M}\right)^2\left(1-{2M \over r} \right)
\left[ -{M(M-r)(2M^2+6Mr-3r^2)\over r^2(2M-r)^2} + {3 \over 2}
\log{\left({r \over r-2M}\right)} \right]\nonumber\\
&&+  3 c_2  \left({r \over M}\right)^2\left(1-{2M \over r} \right).
\nonumber\end{eqnarray}

Equation (23) should be replaced by the following:
 \begin{eqnarray}
&& k_2 = \frac{8C^5}{5}\left(1-2C\right)^2
\left[2+2C\left(y-1\right)-y\right]\times\nonumber\\
&&\bigg\{2C\left(6-3 y+3 C(5y-8)\right)+4C^3\left[13-11y+C(3 y-2)+2
C^2(1+y)\right] \nonumber\\
&& ~ ~
+3(1-2C)^2\left[2-y+2C(y-1)\right]\log\left(1-2C\right)\bigg\}^{-1},\nonumber
\end{eqnarray}

The corrected values for the Love numbers in Table (1) are given in
the table below.

 \clearpage

\begin{deluxetable}{crr}
 \tablecaption{Relativistic Love numbers
$k_2$\label{tbl-1}} \tablewidth{0pt} \tablehead{ \colhead{$n$} &
\colhead{$M/R$} & \colhead{$k_2$}}
\startdata 0.3 & $10^{-5}$ & 0.5511  \\
0.3 & 0.1 & 0.294     \\
0.3 & 0.15 & 0.221\\
0.3 & 0.2 & 0.119\\
0.5 & $10^{-5}$ & 0.4491 \\
0.5 & 0.1 & 0.251 \\
0.5 & 0.15 & 0.173\\
0.5 & 0.2 & 0.095\\
0.5 & 0.25 & 0.0569\\
0.7 & $10^{-5}$ & 0.3626 \\
0.7 & 0.1 & 0.1779 \\
0.7 & 0.15 & 0.1171\\
0.7 & 0.2 & 0.0721\\
0.7 & 0.25 & 0.042\\
1.0 & $10^{-5}$ & 0.2599\\
1.0 & 0.1 & 0.122\\
1.0 & 0.15 & 0.0776\\
1.0 & 0.2 & 0.0459\\
1.0 & 0.2 & 0.0253\\
1.2 & $10^{-5}$ & 0.2062\\
1.2 & 0.1 & 0.0931\\
1.2 & 0.15 & 0.0577\\
1.2 & 0.2 & 0.0327\\
\enddata
\end{deluxetable}

\clearpage

\section{Tidal Love numbers of neutron stars}

\begin{center}
{\bf Tanja Hinderer}\\
{Center for Radiophysics and Space Research, Cornell University,
Ithaca, NY 14853} \\
tph25@cornell.edu
\end{center}

For a variety of fully relativistic polytropic neutron star models
we calculate the star's tidal Love number $k_2$. Most realistic
equations of state for neutron stars can be approximated as a
polytrope with an effective index $n\approx 0.5 - 1.0$. The
equilibrium stellar model is obtained by numerical integration of
the Tolman-Oppenheimer-Volkhov equations. We calculate the linear
$l=2$ static perturbations to the Schwarzschild spacetime following
the method of Thorne and Campolattaro. Combining the perturbed
Einstein equations into a single second order differential equation
for the perturbation to the metric coefficient $g_{tt}$, and
matching the exterior solution to the asymptotic expansion of the
metric in the star's local asymptotic rest frame gives the Love
number. Our results agree well with the Newtonian results in the
weak field limit. The fully relativistic values differ from the
Newtonian values by up to $\sim 24\%$. The Love number is
potentially measurable in gravitational wave signals from
inspiralling binary neutron stars.


\section{Introduction and Motivation}

A key challenge of current astrophysical research is to obtain
information about the equation of state (EoS) of the ultra-dense
nuclear matter making up neutron stars (NSs). The observational
constraints on the internal structure of NSs are weak: the observed
range of NS masses is $M\sim 1.1 - 2.2 M_\odot$ \citep{Lattimer},
and there is no current method to directly measure the radius. Some
estimates using data from X-ray spectroscopy exist, but those are
highly model-dependent  (e. g. \citet{astroph}). Different
theoretical models for the NS internal structure predict, for a
neutron star of mass $M\sim 1.4 M_\odot$, a central density in the
range of $ \rho_c\sim 2-8 \times 10^{14}{\rm g}{\rm cm}^{-3}$ and a
radius in the range of $R\sim 7-16$km \citep{Lattimer}. Potential
observations of pulsars rotating at frequencies above $1400$Hz could
be used to constrain the EoS if the pulsar's mass could also be
measured (e. g. \citet{Bejger}).

Direct and model-independent constraints on the EoS of NSs could be
obtained from gravitational wave observations. Coalescing binary
neutron stars are one of the most important sources for ground-based
gravitational wave detectors \citep{Cutler:2002me}. LIGO
observations have established upper limits on the coalescence rate
per comoving volume \citep{Abbott:2007xi}, and at design sensitivity
LIGO II is expected to detect inspirals at a rate of $\sim 2/$day
\citep{2004ApJ...601...L179}.

In the early, low frequency part of the inspiral ($f\leq 100$Hz,
where $f$ is the gravitational wave frequency), the waveform's phase
evolution is dominated by the point-mass dynamics and finite-size
effects are only a small correction. Toward the end of the inspiral
the internal degrees of freedom of the bodies start to appreciably
influence the signal, and there have been many investigations of how
well the EoS can be constrained using the last several orbits and
merger, including constraints from the gravitational wave energy
spectrum \citep{faber} and from the NS tidal disruption signal for
NS-black hole binaries \citep{vallisneri}. Several numerical
simulations of the hydrodynamics of NS-NS mergers have studied the
dependence of the gravitational wave spectrum on the radius and EoS
(see, e.g. \citet{baumgarte} and references therein). However,
trying to extract EoS information from this late time regime
presents several difficulties: (i) the highly complex behavior
requires solving the full nonlinear equations of general relativity
together with relativistic hydrodynamics; (ii) the signal depends on
unknown quantities such as the spins and angular momentum
distribution inside the stars, and (iii) the signals from the
hydrodynamic merger are outside of LIGO's most sensitive band.

During the early regime of the inspiral the signal is very clean and
the influence of tidal effects is only a small correction to the
waveform's phase. However, signal detection is based on matched
filtering, i. e. integrating the measured waveform against
theoretical templates, where the requirement on the templates is
that the phasing remain accurate to $\sim 1$ cycle over the
inspiral. If the accumulated phase shift due to the tidal
corrections becomes of order unity or larger, it could corrupt the
detection of NS-NS signals or alternatively, detecting a phase
perturbation could give information about the NS structure. This has
motivated several analytical and numerical investigations of tidal
effects in NS binaries
\citep{bc,ks,kochanek,ts,mw,shibata,gualteri,pons,berti}. The
influence of the internal structure on the gravitational wave phase
in this early regime of the inspiral is characterized by a single
parameter, namely the ratio $\lambda$ of the induced
 quadrupole to the perturbing tidal field. This ratio $\lambda$ is related to
the star's tidal Love number $k_2$ by $k_2 = 3G \lambda R^{-5}/2$,
where $R$ is the star's radius. \citet{tidal} have shown that for an
inspiral of two non-spinning $1.4M_{\odot}$ NSs at a distance of 50
Mpc, LIGO II detectors will be able to constrain $\lambda$ to
$\lambda \leqslant 2.01 \times 10^{37} {\rm g}{\,}{\rm cm}^2{\rm
s}^2 $ with $90\%$ confidence. This number is an upper limit on
$\lambda$ in the case that no tidal phase shift is observed. The
corresponding constraint on radius would be $R \leqslant 13.6
{\,}{\rm km} {\;}\left(15.3{\,} {\rm km} \right)$ for a $n=0.5$
$\left(n=1.0\right)$ fully relativistic polytrope, for
$1.4M_{\odot}$ NSs \citep{tidal}.

 Because neutron stars are compact
objects with strong internal gravity, their Love numbers could be
very different from those for Newtonian stars that have been
computed previously, e. g. by \citet{bo}.

 Knowledge of Love number values could also be useful for
comparing different numerical simulations of NS binary inspiral by
focusing on models with the same masses and values of $\lambda$.

In \citet{tidal}, the $l=2$ tidal Love numbers for fully
relativistic neutron star models with polytropic pressure-density
relation $P=K\rho^{1+1/n}$, where $K$ and $n$ are constants, were
computed for the first time. The present paper will give details of
this computation. Using polytropes allows us to explore a wide range
of stellar models, since most realistic models can be reasonably
approximated as a polytrope with an effective index in the range
$n\sim 0.5-1.0$ \cite{Lattimer}. Our prescription for computing
$\lambda$ is valid for an arbitrary pressure-density relation and
not restricted to polytropes. In Sec. \ref{sec:lovedef}, we start by
defining $\lambda$ in the fully relativistic context in terms of
coefficients in an asymptotic expansion of the metric in the star's
local asymptotic rest frame and discuss the extent to which it is
uniquely defined. In Sec. \ref{sec:lovecalc}, we discuss our method
of calculating $\lambda$, which is based on static linearized
perturbations of the equilibrium configuration in the Regge-Wheeler
gauge as in \citet{tc}. Section \ref{sec:results} contains the
results of the numerical computations together with a discussion.
Unless otherwise specified, we use units in which $c=G=1$.

\section{Definition of the Love number}
\label{sec:lovedef} Consider a static, spherically symmetric star of
mass $M$ placed in a static external quadrupolar tidal field  ${\cal
E}_{ij}.$ The star will develop in response a quadrupole moment
$Q_{ij}$ {\footnote{ The induced quadrupolar deformation of the star
can be described in terms of the star's $l=2$ mode eigenfunctions of
oscillation.}. In the star's local asymptotic rest frame
(asymptotically mass centered Cartesian coordinates) at large $r$
the metric coefficient $g_{tt}$ is given by \citep{Thorne1998}:
\begin{eqnarray} \frac{\left(1-g_{tt} \right)}{2} &=& - {M \over r}- {3Q_{ij}
\over 2 r^3} \left( n^i n^j -
\frac{1}{3}\delta^{ij} \right)+O\left(\frac{1}{r^3}\right)\nonumber\\
& &+\frac{1}{2}{\cal E}_{ij} x^i x^j + O\left(r^3\right),
\label{Qdef} \end{eqnarray} where $n^i=x^i/r;$ this expansion
defines ${\cal E}_{ij}$ \footnote{The $l=2$ tidal moment can be
related to a component of the Riemann tensor
$R_{\alpha\beta\gamma\delta}$ of the external pieces of the metric
in Fermi normal coordinates at $r=0$ as ${\cal E}_{ij}=R_{0i0j}$
(see \citet{mtw}).} and $Q_{ij}$. In the Newtonian limit, $Q_{ij}$
is related to the density perturbation $\delta \rho$ by
\begin{equation} Q_{ij}=\int d^3 x \delta\rho({\bf x}) \left( x_i x_j -
\frac{1}{3}r^2
  \delta_{ij} \right),\label{qnewtonian}\end{equation}
and ${\cal E}_{ij}$ is given in terms of the external gravitational
potential $\Phi_{\rm ext}$ as \begin{equation} {\cal
E}_{ij}=\frac{\partial^2 \Phi_{\rm ext}}{\partial x^i\partial
x^j}.\label{enewtonian}\end{equation} We are interested in
applications to fully relativistic stars, which requires going
beyond Newtonian physics. In the strong field case, Eqs.
(\ref{qnewtonian}) and (\ref{enewtonian}) are no longer valid but
the expansion of the metric (\ref{Qdef}) still holds in the
asymptotically flat region and serves to define the moments $Q_{ij}$
and ${\cal E}_{ij}$.

We briefly review here the extent to which these moments are
uniquely defined since there are considerable coordinate ambiguities
in performing asymptotic expansions of the metric. For an isolated
body in a static situation these moments are uniquely defined:
${\cal E}_{ij}$ and $Q_{ij}$ are the coordinate independent moments
 defined by \citet{geroch} and \citet{hansen} for stationary, asymptotically flat
spacetimes in terms of certain combinations of the derivatives of
the norm and twist of the timelike Killing vector at spatial
infinity. In the case of an isolated object in a dynamical
situation, there are ambiguities related to gravitational radiation,
for example angular momentum is not uniquely defined \citep{wald}.
For the application to the adiabatic part of a NS binary inspiral,
we are interested in the case of a non-isolated object in a
quasi-static situation. In this case there are still ambiguities
(related to the choice of coordinates) but their magnitudes can be
estimated \citep{thornehartle} and are at a high post-Newtonian
order and therefore can be neglected. We are also interested in (i)
working to linear order in ${\cal E}_{ij}$ and (ii) in the limit
where the source of ${\cal E}_{ij}$ is very far away. In this limit
the ambiguities disappear.

To linear order in ${\cal E}_{ij}$, the induced quadrupole will be
of the form \begin{equation} Q_{ij} =- \lambda {\cal E}_{ij}.
\label{Lovedef} \end{equation} Here $\lambda$ is a constant which is
related to the $l=2$ tidal Love number (apsidal constant) $k_2$  by
\citep{tidal} \begin{equation} k_2 = \frac{3}{2}G \lambda
R^{-5}.\label{k2def}\end{equation} Note the difference in
terminology: in \citet{tidal}, $\lambda$ was called the Love number,
whereas in this paper, we reserve that name for the dimensionless
quantity $k_2$.

The tensor multipole moments $Q_{ij}$ and ${\cal E}_{ij}$ can be
decomposed as \begin{equation} {\cal E}_{ij}=\sum^2_{m=-2}{\cal E}_m
{\cal Y}^{2m}_{ij},\label{emdef}\end{equation} and \begin{equation}
{Q}_{ij}=\sum^2_{m=-2}{Q}_m {\cal
Y}^{2m}_{ij},\label{qmdef}\end{equation} where the symmetric
traceless tensors ${\cal Y}^{2m}_{ij}$ are defined by
\citep{Thorne1980} \begin{equation} Y_{2m}(\theta, \varphi)={\cal
Y}^{2m}_{ij}n^i n^j\end{equation} with ${\bf
n}=(\sin\theta\cos\varphi,\sin\theta\sin\varphi,\cos \theta)$. Thus,
the relation (\ref{Lovedef}) can be written as \begin{equation}
Q_{m} =- \lambda {\cal E}_{m}. \label{Lovedef1} \end{equation}
Without loss of generality, we can assume that only one ${\cal E}_m$
is nonvanishing, this is sufficient to compute $\lambda$.

\section{Calculation of the Love number}
\label{sec:lovecalc}
\subsection{Equilibrium configuration} \label{subsecequil} The
geometry of spacetime of a spherical, static star can be described
by the line element \citep{mtw}
\begin{equation}
ds_0^2=g^{(0)}_{\alpha \beta}dx^\alpha
dx^\beta=-e^{\nu(r)}dt^2+e^{\lambda(r)}dr^2+r^2
\left(d\theta^2+\sin^2\theta d \phi^2 \right).
\end{equation} The star's stress-energy
tensor is given by \begin{equation} T_{\alpha
\beta}=\left(\rho+p\right)u_\alpha u_\beta+pg^{(0)}_{\alpha
\beta},\label{tab}\end{equation} where
$\vec{u}=e^{-\nu/2}\partial_t$ is the fluid's four-velocity and
$\rho$ and $p$ are the density and pressure. Numerical integration
of the Tolman-Oppenheimer-Volkhov equations (see e.g. \citet{mtw})
for neutron star models with a polytropic pressure-density relation
\begin{equation} P=K\rho^{1+1/n},\end{equation} where $K$ is a constant and $n$ is the
polytropic index, gives the equilibrium stellar model with radius
$R$ and total mass $M=m(R)$.

\subsection{Static linearized perturbations due to an external
tidal field } \label{subsecpert} We examine the behavior of the
equilibrium configuration under linearized perturbations due to an
external quadrupolar tidal field following the method of \citet{tc}.
The full metric of the spacetime is given by \begin{equation}
g_{\alpha\beta}=g^{(0)}_{\alpha \beta} +h_{\alpha
\beta},\label{gtotal}\end{equation} where $h_{\alpha\beta}$ is a
linearized metric perturbation. We analyze the angular dependence of
the components of $h_{\alpha\beta}$ into spherical harmonics as in
\citet{rw}. We restrict our analysis to the $l=2$, static,
even-parity perturbations in the Regge-Wheeler gauge \citep{rw}.
With these specializations, $h_{\alpha\beta}$ can be written as
\citep{rw, tc}:
\begin{equation} h_{\alpha \beta} ={\rm diag}
\left[e^{-\nu(r)}H_0(r), ~ e^{\lambda(r)} H_2(r), ~ r^2 K(r), ~
r^2 \sin^2\theta K(r)\right] Y_{2m}(\theta, \varphi)\label{hrw}.
\end{equation}
The nonvanishing components of the perturbations of the
stress-energy tensor (\ref{tab}) are $\delta T^0_0=-\delta
\rho=-(dp/d\rho)^{-1}\delta p$ and $\delta T^i_i=\delta p$. We
insert this and the metric metric perturbation (\ref{hrw}) into the
the linearized Einstein equation $\delta G_{\alpha}^{\beta} = 8 \pi
\delta T_{\alpha}^{\beta}$ and combine various components. From
$\delta G^\theta_\theta-\delta G^\phi_\phi=0$ it follows that that
$H_2=H_0\equiv H$, then $\delta G^r_\theta=0$ relates $K'$ to $H$,
and after using $\delta G_\theta ^\theta+\delta G_\phi^\phi=16\pi
\delta p$ to eliminate $\delta p$, we finally subtract the $r-r$
component of the Einstein equation from the $t-t$ component to
obtain the following differential equation for $H_0\equiv H$ (for
$l= 2$):
\begin{eqnarray}
&&H{''}+H{'} \left[{2 \over r} + e^{\lambda} \left( {2m(r)\over
r^2}
+ 4 \pi r \left(p-\rho\right)\right) \right]\nonumber\\
&&+H\left[ -{6 e^{\lambda} \over r^2 } + 4 \pi e^{\lambda}\left( 5
\rho + 9 p +
 {\rho + p \over \left(dp/d\rho\right)} \right)
 - \nu{'}^2 \right]=0,
\label{Hdiffeq}\end{eqnarray} where the prime denotes $d/dr$. The
boundary conditions for Eq. (\ref{Hdiffeq}) can be obtained as
follows. Requiring regularity of $H$ at $r=0$ and solving for $H$
near $r=0$ yields \begin{equation} H(r)=a_0
r^2\left[1-\frac{2\pi}{7}\left(5\rho(0)+9p(0)+\frac{\rho(0)+p(0)}{(dp/d\rho)(0)}\right)r^2+
O(r^3)\right], \label{horigin}\end{equation} where $a_0$ is a
constant. To single out a unique solution from this one-parameter
family of solutions parameterized by $a_0$, we use the continuity of
$H(r)$ and its derivative across $r=R$. Outside the star, Eq.
(\ref{Hdiffeq}) reduces to
\begin{equation}
H{''} + \left( {2\over r} - \lambda{'}\right) H{'} -\left( {6
e^{\lambda} \over r^2 } + \lambda{'}^2 \right) H= 0,
\label{Hdiffeqouts1}
\end{equation}
and changing variables to $x =( r/M -1)$ as in \citet{tc} transforms
Eq. (\ref{Hdiffeqouts1}) to a form of the associated Legendre
equation with $l=m=2$:
\begin{equation}
\left(x^2 -1 \right) H{''} + 2 x H{'} - \left( 6 + {4\over x^2 -1 }
\right)H=0.\label{houtsideeq}
\end{equation}
The general solution to Eq. (\ref{houtsideeq}) in terms of the
associated Legendre functions $Q_2{\,}^2(x)$ and $P_2{\,}^2(x)$ is
given by
\begin{equation}H = c_1 Q_2{\,}^2 \left( {r \over M}-1 \right) + c_2
P_2{\,}^2\left( {r \over M}-1 \right),
\end{equation}
where $c_1$ and $c_2$ are coefficients to be determined.
Substituting the expressions for $Q_2{\,}^2(x)$ and $P_2{\,}^2(x)$
from \citet{as} yields for the exterior solution \begin{eqnarray}
H&=&c_1 \left({r \over M}\right)^2\left(1-{2M \over r} \right)
\left[ -{M(M-r)(2M^2+6Mr-3r^2)\over r^2(2M-r)^2} + {3 \over 2}
\log{\left({r \over r-2M}\right)} \right]\nonumber\\
&&+  3 c_2  \left({r \over M}\right)^2\left(1-{2M \over r} \right).
\label{Hout}\end{eqnarray} The asymptotic behavior of the solution
(\ref{Hout}) at large $r$ is
\begin{equation} H = {8 \over 5} \left({M \over
r}\right)^3 c_1+O\left(\left(\frac{M}{r}\right)^4\right) + 3\left({r
\over M}\right)^2
c_2+O\left(\left(\frac{r}{M}\right)\right)\label{asympH},\end{equation}
where the coefficients $c_1$ and $c_2$ are determined by matching
the asymptotic solution (\ref{asympH}) to the expansion (\ref{Qdef})
and using Eq. (\ref{Lovedef1}):
\begin{equation} c_1 = {15 \over 8}{1 \over M^3} \lambda {\cal{E}},
{\;}{\;}{\;}{\;}{\;}c_2 = {1 \over 3} M^2
{\cal{E}}.\label{coefficients}\end{equation} We now solve for
$\lambda$ in terms of $H$ and its derivative at the star's surface
$r=R$ using Eqs. (\ref{coefficients}) and (\ref{Hout}), and use the
relation (\ref{k2def}) to obtain the expression:
 \begin{eqnarray} \label{k2expr}
&& k_2 = \frac{8C^5}{5}\left(1-2C\right)^2
\left[2+2C\left(y-1\right)-y\right]\times\\
&&\bigg\{2C\left(6-3 y+3 C(5y-8)\right)+4C^3\left[13-11y+C(3 y-2)+2
C^2(1+y)\right] \nonumber\\
&& ~ ~
+3(1-2C)^2\left[2-y+2C(y-1)\right]\log\left(1-2C\right)\bigg\}^{-1},\nonumber
\end{eqnarray} where we have defined the star's compactness parameter $C\equiv
M/R$ and the quantity $y\equiv RH{'}(R)/ H(R)$, which is obtained by
integrating Eq. (\ref{Hdiffeq}) outwards in the region $0<r<R$.

\subsection{Newtonian limit}

The first term in the expansion of the expression (\ref{k2expr}) in
$M/R$ reproduces the Newtonian result:
\begin{equation} k_2^N=\frac{1}{2 }\left(\frac{2- y}{ y + 3
}\right),\label{newtk2}\end{equation} where the superscript $N$
denotes "Newtonian". In the Newtonian limit, the differential
equation (\ref{Hdiffeq}) inside the star becomes \begin{equation}
H''+\frac{2}{r}H'+\left(\frac{4\pi\rho}{dp/d\rho}-\frac{6}{r^2}\right)H=0.\label{Newtoniandiff}
\end{equation} For a polytropic index of $n=1$, Eq. (\ref{Newtoniandiff}) can
be transformed to a Bessel equation with the solution that is
regular at $r=0$ given by $ H=A \sqrt{{r/R}} ~ J_{5/2}(\pi r/ R)$,
where $A$ is a constant. At $r=R,$ we thus have $y=RH{'}/ H =(\pi^2
- 9 )/3,$ and from Eq. (\ref{k2expr}) it follows that
\begin{equation} k_2^N(n=1)=\left( -{1 \over 2} + {15 \over 2 \pi^2}\right) \approx
0.25991,\end{equation} which agrees with the known result of
\citet{bo}.

\section{Results and Discussion}
\label{sec:results}

The range of dimensionless Love numbers $k_2$ obtained by numerical
integration of Eq. (\ref{k2expr}) is shown in Fig. \ref{dimlam} as a
function of $M/R$ and $n$ for a variety of different neutron star
models, and representative values are given in Table \ref{tbl-1}.
These values can be approximated to an accuracy of $\sim 6\%$ in the
range $0.5\leq n\leq 1.0$ and $0.1\leq (M/R)\leq 0.24$ by the
fitting formula
\begin{equation} k_2\approx \frac{3}{2}
\left(-0.41+\frac{0.56}{n^{0.33}}\right)\left(\frac{M}{R}\right)^{-0.003}.\end{equation}

Both Fig. \ref{dimlam} and Table \ref{tbl-1} illustrate that the
dimensionless Love numbers $k_2$ depend more strongly on the
polytropic index $n$ than on the compactness $C=M/R$.
\footnote{Note, however, that LIGO measurements will yield the
combination $k_2 R^5$ and therefore will be more sensitive to the
compactness than the polytropic index.} This is expected since the
weak field, Newtonian values $k_2^N$ given by Eq. (\ref{newtk2})
just depend on $n$ (through the dependence on $y$). The additional
dependence on the compactness for the Love numbers $k_2$ in Eq.
(\ref{k2expr}) is a relativistic correction to this. For $M/R\sim
10^{-5}$ our results for $k_2$ agree well with the Newtonian results
of \citet{bo}. Figure \ref{diff} shows the percent difference
$(k_2^N-k_2)/k_2$ between the relativistic and Newtonian
dimensionless Love numbers. As can be seen from the figure, the
relativistic values are lower than the Newtonian ones for higher
values of $n$. This can be explained by the fact that the Love
number encodes information about the degree of central condensation
of the star. Stars with a higher the polytropic index $n$ are more
centrally condensed and therefore have a smaller response to a tidal
field, resulting in a smaller Love number.

Some estimates of the masses and radii of neutron stars, given in
Table \ref{tbl-2}, have been inferred from X-ray observations
\citep{nature,astroph} using the information from three measured
quantities: the Eddington luminosity, the surface redshift of
spectral lines, and the quiescent X-ray flux. The range of the
numbers $\lambda$ for these stars is shown in Fig. \ref{xray}. LIGO
II detectors will be able to establish a $90\%$ confidence upper
limit of $\lambda \leqslant 2.01 \times 10^{37} {\rm g}{\,}{\rm
cm}^2{\rm s}^2 $ for an inspiral of two nonspinning $1.4 M_\odot$
NSs at a distance of 50 Mpc in the case that no tidal phase shift is
observed \citep{tidal}.

\acknowledgments

The author thanks \'Eanna Flanagan for valuable discussions and
comments.

\clearpage

\begin{figure}
\epsscale{.70} \plotone{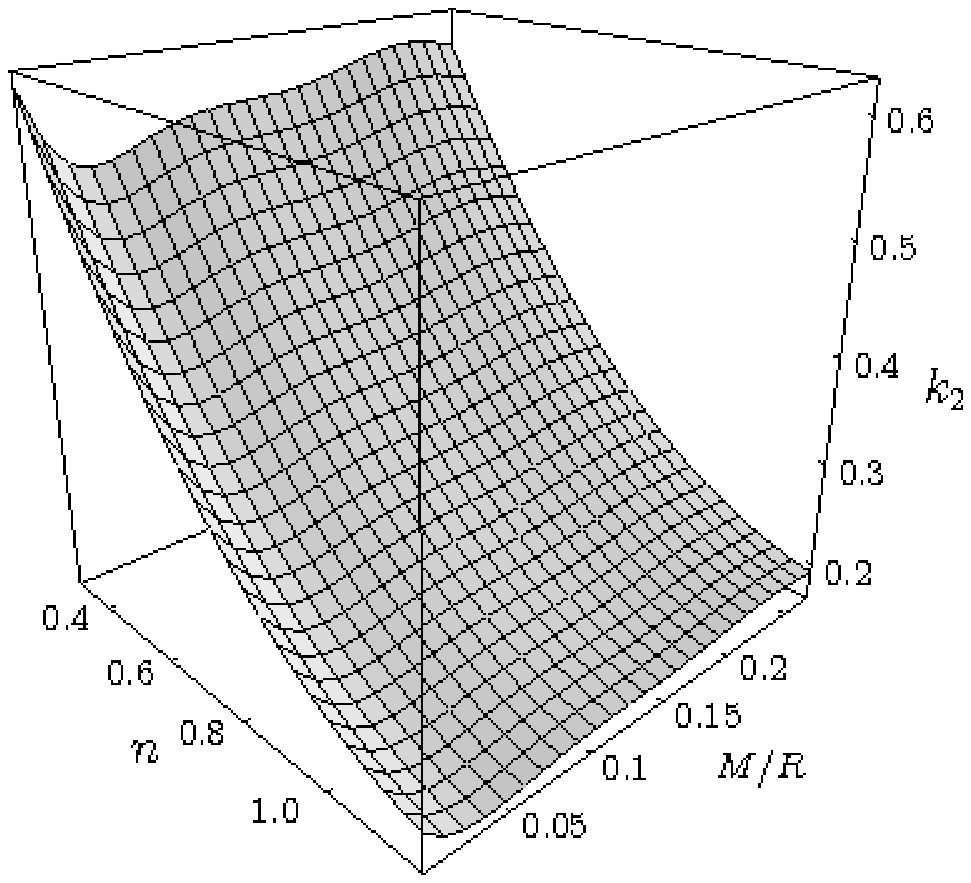} \caption{The relativistic Love
numbers $k_2$. } \label{dimlam}
\end{figure}

\begin{figure}
\epsscale{.80} \plotone{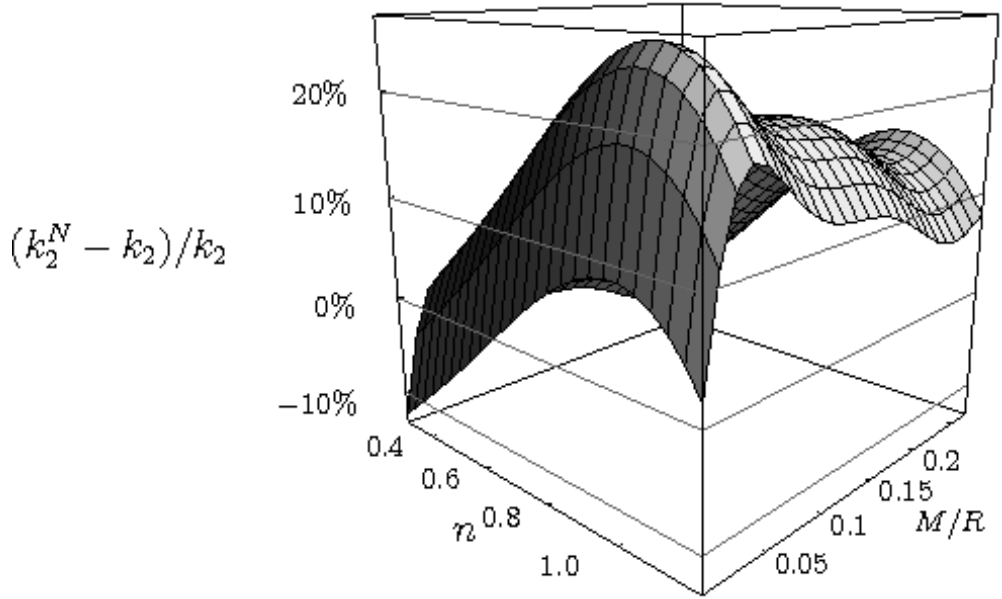} \caption{The difference in percent
between the relativistic dimensionless Love numbers $k_2$ and the
Newtonian values $k_2^N$.}\label{diff}
\end{figure}

\begin{figure}
\epsscale{.80} \plotone{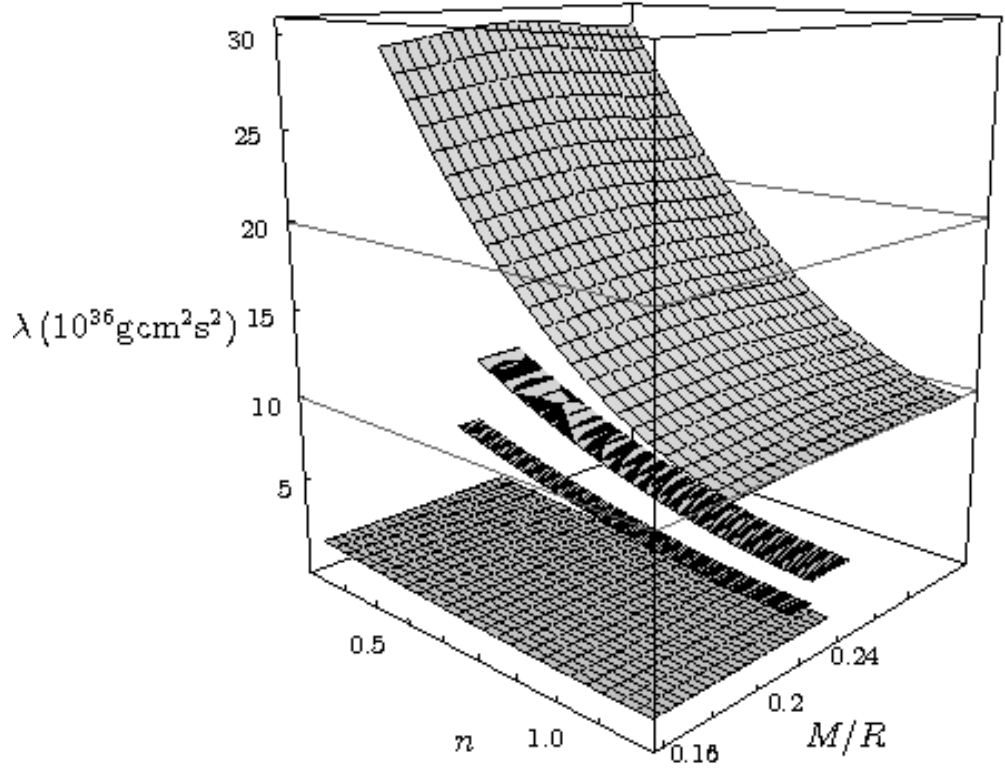} \caption{ The range of Love numbers
for the estimated NS parameters from X-ray observations. Top to
bottom sheets: EXO0748-676, $\omega$Cen, M 13, NGC 2808. For an
inspiral of two $1.4M_{\odot}$ NSs at a distance of 50 Mpc, LIGO II
detectors will be able to constrain $\lambda$ to $\lambda \leqslant
20.1 \times 10^{36} {\rm g}{\,}{\rm cm}^2{\rm s}^2 $ with $90\%$
confidence \citep{tidal}. }\label{xray}
\end{figure}

\clearpage

\begin{deluxetable}{crr}
 \tablecaption{Relativistic Love numbers
$k_2$\label{tbl-1}} \tablewidth{0pt} \tablehead{ \colhead{$n$} &
\colhead{$M/R$} & \colhead{$k_2$}}
\startdata 0.3 & $10^{-5}$ & 0.5511  \\
0.3 & 0.1 & 0.294     \\
0.3 & 0.15 & 0.221\\
0.3 & 0.2 & 0.119\\
0.5 & $10^{-5}$ & 0.4491 \\
0.5 & 0.1 & 0.251 \\
0.5 & 0.15 & 0.173\\
0.5 & 0.2 & 0.095\\
0.5 & 0.25 & 0.0569\\
0.7 & $10^{-5}$ & 0.3626 \\
0.7 & 0.1 & 0.1779 \\
0.7 & 0.15 & 0.1171\\
0.7 & 0.2 & 0.0721\\
0.7 & 0.25 & 0.042\\
1.0 & $10^{-5}$ & 0.2599\\
1.0 & 0.1 & 0.122\\
1.0 & 0.15 & 0.0776\\
1.0 & 0.2 & 0.0459\\
1.0 & 0.2 & 0.0253\\
1.2 & $10^{-5}$ & 0.2062\\
1.2 & 0.1 & 0.0931\\
1.2 & 0.15 & 0.0577\\
1.2 & 0.2 & 0.0327\\
\enddata
\end{deluxetable}

\clearpage

\begin{table}
\begin{center}
\caption{Estimated neutron star parameters from X-ray
observations\label{tbl-2}}
\begin{tabular}{cccc}
\tableline\tableline Cluster / object & $M (M_\sun)$ & $R(\rm km)$ &
$M/R$ \\
\tableline
$\omega$ Cen \tablenotemark{a} & $1.61\pm 0.15$ & $10.99\pm 0.71$ & $0.18\pm 0.04$ \\
M 13 \tablenotemark{a}& $1.36\pm 0.04$ & $9.89\pm 0.08$ & $0.2$ \\
NGC 2808 \tablenotemark{a}&$0.84\pm 0.12$ & $7.34\pm 0.96$ & $0.22\pm 0.01$ \\
EXO 0748-676 \tablenotemark{b}&$\geq 2.1\pm 0.28$ & $\geq 13.8\pm 1.8$ & $0.2256$ \\
\tableline
\end{tabular}
\tablenotetext{a}{The parameters for these stars are the averages
from the best fit values of the data in \citet{astroph} for their
three different spectral fits. The errors given here reflect only
the deviations among the best fit values for the fits.}
\tablenotetext{b}{The values are taken from \citet{nature}.}
\tablecomments{Parameters used to generate Fig. (\ref{xray}).}
\end{center}
\end{table}

\end{document}